# Detecting fog where satellite systems are limited

Noam David[†]

†The school of Civil and Environmental Engineering, Cornell University, Ithaca, NY 14853, United States.

E mail: nd363@cornell.edu

**ABSTRACT**

Surface level hydrometeors, and specifically fog, cause attenuation of the signal level measured by microwave communication networks. These networks operate at frequencies of tens of GHz and form the data transport infrastructure between cellular base stations. Thus, wireless communication links are, effectively, an existing fog monitoring facility. Operational costs are minimal, since the network is already deployed in the field, and measurements are already stored by many of the cellular providers for quality assurance purposes.

This study shows the potential of microwave communication networks for the detection of fog in challenging conditions where satellite systems are often limited. In a first event, the ability to detect fog using commercial microwave links, at a time when the satellite cannot detect the phenomena due to high level cloud cover obscuring the ground level fog, is demonstrated. The ability of the microwave system to rule out ground level fog at times when the satellite detects a low–lying stratus cloud, but cannot identify whether it is adjacent to the ground or at higher elevations above it, is demonstrated in a second event. The results indicate the operative potential of this technique to assist end users in reliable decision making that have highly important economic and safety implications.

# 1 INTRODUCTION

Fog is defined as a state where water droplets, suspended near ground level, limit visibility to less than 1 kilometer (Glickman, 2000). This phenomenon is a major hazard in areas of intense traffic including airports, piers and highways, due to the visibility impairment it creates (Gultepe et al., 2009; Ashley et al., 2015). According to Rosenfeld (1996), for example, fog-related road accidents caused the death of twice the number of people, between 1982 and 1991, than those who died from the extreme phenomena of hurricanes, lightning and tornadoes combined. Another example is taken from Illinois where, excluding the city of Chicago, approximately 4,000 accidents and 30 fatalities occur annually in ground transportation, due to foggy conditions, according to Westcott (2007).  Musk (1991) finds increased accident rates, with a higher percentage of multiple collisions than normal, in cases of thick fog.  Where aviation is concerned, fog may be a central cause of fatal accidents. For instance, according to a study held by the National Transportation Safety Board between 1995 and 2000, 63% of all weather-related aviation disasters occurred during periods with low ceilings, clouds and fog (Pearson, 2002; Cox, 2007). The attempts to avert the dangers posed to air traffic from the occurrence of dense fog cause flight cancellations and airport delays that can lead to vast financial damages (Kutty et al., 2018). Thus, for example, as Gultepe et al. (2009) reviews, between December 20[th] and 23[rd], 2006, 175,000 passengers were forced to miss their flights at seven UK airports due to intense fog existing in the area (Milmo, 2007). Preliminary estimates assessed the damages at a minimum of £25M (Gadher and Baird, 2007).

Forecasting and monitoring of fog is, therefore, crucial for public safety and is of great economic value.

The available instruments of fog monitoring include satellites (e.g. Lensky and Rosenfeld, 2008; Dey, 2017), and ground based instruments such as transmissometers, visibility sensors, and human observers (WMO, 2008; David et al., 2013). Satellites provide information with high spatial resolution, but have difficulty, at times, in detecting ground level fog, due to, for example, high or mid-level cloud layers that may obscure the fog from the satellite vantage point (Gultepe et al., 2007a). Furthermore, satellites may find it difficult to distinguish between fog lying in very close proximity to the ground, and a stratus cloud at an altitude of tens of meters AGL, an elevation that does not endanger drivers, for instance. Standard visibility sensors derive a reliable point measurement, but with low spatial representativity (Gultepe et al., 2007b). Transmissometers, which comprise a transmitter and receiving unit, typically deployed a few tens of meters apart, provide reliable visibility estimates, but the equipment cost is very high, and as a result, they are only installed in very specific sites, such as airports, where they are absolutely necessary.  Naturally, human observers detect fog and estimate the resulting visibility most reliably, when compared to the other instruments, but this way of detecting the phenomenon is not automatic, and the visibility estimates are subjective, and affected by psycho-physical factors, and based on each observer's different visual acuity. Moreover, the different monitoring instruments require continuous maintenance, and due to the high costs involved, it is not feasible to deploy them over a wide swath of terrain.

Thus, at this point in time, due to economic and practical limitations, there is no tool that can reliably provide large-scale observations of ground level fog with low costs.

A unique way of generating the required information at minimum cost is using measurements from cellular network communication infrastructure.

Cellular communication networks operate on the principle where each mobile phone located in a specific sector, communicates with the network base station that is nearest to it. This base station transmits the data to the next base station, and then to the next, until the transmitted signal reaches the base station situated near the second end user, located at a different point in space. A cost effective method for transferring data between base stations, which are fixed in their locations, utilizes microwave radiation at frequencies of tens of GHz, transmitted very close to ground level, at elevations of tens of meters. Figure 1 shows an example of a microwave antenna mast.

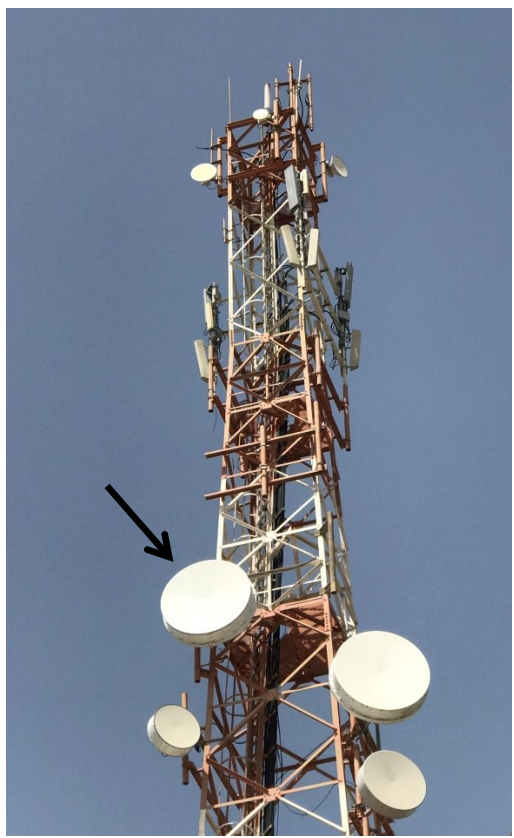

Figure 1. Communication mast. The drum like microwave antennas (indicated by the arrow) are located in close proximity to the surface and can be used for sensing fog (Photo credit: N. David).

Since different atmospheric conditions affect the Received Signal Level (RSL) between these base stations, it can be utilized as an environmental monitoring system, as initially demonstrated for rain observations (Messer et al., 2006; Leijense et al., 2007) and mapping (e.g. Goldshtein et al., 2009; Rayitsfeld et al., 2012; Overeem et al., 2013; Liberman et al., 2014). The potential for monitoring atmospheric water vapor (David et al., 2009; 2011; Chwala et al., 2013), and for dew detection based on the measurements from these networks were also shown (Harel et al., 2015). Recently, David et al. (2013) presented the feasibility for monitoring dense fog using commercial microwave links, and pointed out the potential of future operating frequencies for monitoring the phenomena at higher resolution (David et al., 2015). Temperature inversions suppress vertical atmospheric movements and create ideal conditions for increased air pollution and fog. On the other hand, under temperature inversion conditions, anomalous microwave propagation conditions may be created and recent studies have indicated the potential to detect air pollution and fog based on this principle (David and Gao, 2016; 2018).

The data required for carrying out the environmental observations using commercial microwave links is the received signal level between the base stations of the communication network itself, and therefore, doesn't involve end user privacy issues of any kind.

This study shows the potential of the technology to provide large-spatial measurements of fog during periods where observation by proprietary satellite systems is not possible due to cloud cover at altitude, or at times where the satellite is detecting stratus clouds that, in practice, lie at altitudes above ground level.

The concept is demonstrated using standard microwave data that is routinely collected by many cellular operators, during normal system operation, for quality of service needs.

## 2 SCIENTIFIC BACKGROUND AND METHOD

### 2.1 Fog induced attenuation in the microwave region

The For frequencies of up to 200 GHz, a Rayleigh approximation model is valid and can be used for the calculation of fog induced attenuation (Rec. ITU-R P.840-4, 2009; David et al., 2013):

$$(1) \quad \Gamma_f = \Theta(f,T) \cdot LWC \qquad \text{(dB/km)}$$

Where:

$\Gamma_f$ – Attenuation due to fog (dB/km).

LWC- Fog Liquid Water Content (gr/m$^3$).

θ (f, T)-  Specific attenuation coefficient ((dB/km)/(g/m$^3$)), where $f$ (GHz) and $T$ (K) are the link operating frequency and the temperature, respectively.

Eq. (2) and (3) detail the procedure by which θ can be calculated:

$$(2) \quad \Theta = \frac{0.819 f}{\varepsilon''(1 + \mu^2)} \qquad \text{(dB/km)/(g/m}^3)$$

Where μ is:   (3)   $\quad \mu = \dfrac{2 + \varepsilon'}{\varepsilon''}$

The complex dielectric permittivity of water is denoted by ε. The complete expression of this parameter is detailed in literature (Rec. ITU-R P.840-4, 2009). Figure (2), which was generated using Eq. (1), presents the theoretical fog induced attenuation asa function of microwave frequencies. Notably, for a given Liquid

Water Content (LWC) and temperature, fog induced attenuation increases with an increase in frequency, and as frequency increases, the sensitivity for monitoring the phenomenon increases (David et al., 2015).

Beyond fog itself, atmospheric gases also cause signal attenuation, hence the importance of examining their effects on the system.

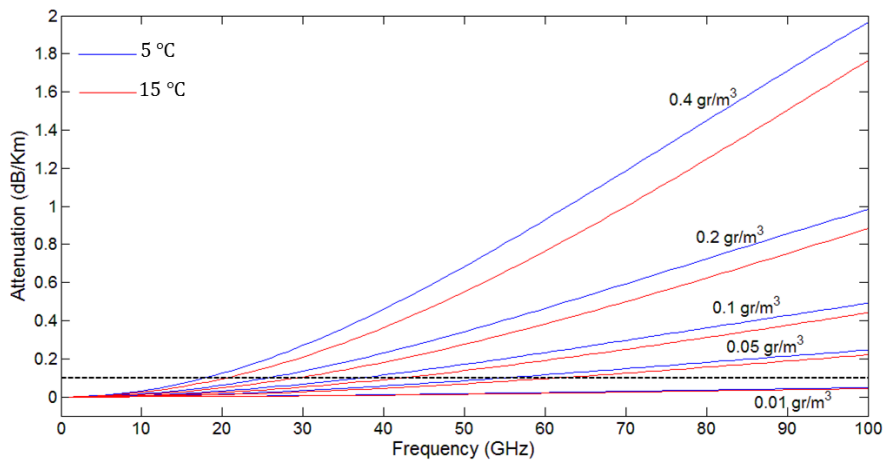

Figure 2. Transmission loss due to fog-LWC as a function of microwave frequencies up to 100 GHz. The calculation was carried out for temperatures of 5 °C (blue curves) and 15 °C (red curves) and for LWC ranging from 0.01 to 0.4 gr/m$^3$. The horizontal dashed line represents the quantization error of the microwave system used in this research (0.1 dB).

## 2.2 Attenuation by atmospheric gases

Evaluating the specific attenuation due to dry air and water vapor, for any value of pressure, temperature and humidity, and at frequencies up to 1000 GHz can be achieved by summation of the different attenuating factors - the individual resonance lines from oxygen and water vapor, as well as the additional small factors of pressure induced nitrogen attenuation above 100 GHz, the non-resonant

Debye spectrum of oxygen below 10 GHz, and a wet continuum to account for the experimentally found excess water vapor absorption (Rec. ITU-R P.676-6, 2005; David et al., 2009). Equation (4) shows the specific gaseous attenuation, $\gamma$ (dB/km):

$$\gamma = \gamma_o + \gamma_w = 0.1820 f \left( N''(f, p, T, e) \right) \qquad \text{(dB/km)} \qquad (4)$$

Where $\gamma_0$ and $\gamma_w$ are the dry air (oxygen, pressure-induced nitrogen and non-resonant Debye attenuation) and water vapor induced attenuations, respectively. $N''$ (N-units) is the imaginary part of the complex refractivity, a function of the frequency- f (GHz), pressure- p (hPa), temperature- T($°C$) and the water vapor partial pressure- e (hPa).

Full details of the functions and parameters in Equation (4) can be found in the literature (Rec. ITU-R P.676-6, 2005).

The continuous curves in Figure (3) show the attenuation calculated according to Equation (4) for dry air (in red), and water vapor (in blue) between 1 and 100 GHz, and with standard conditions of 1013 millibar, 15°C, and water vapor density of 10 gr/m$^3$ . The oxygen molecule, and water vapor absorption bands can be seen around 60 GHz and 22.2 GHz respectively. The dotted lines (in blue) represent the dynamic range of induced attenuation, based on the minimum and maximum values characteristic of the absolute humidity in the areas where the measurements for this study were taken (David et al., 2011), that is, between 5 and 20  gr/m$^3$, respectively.

Additionally, in order to calculate the attenuation range, typical fluctuations of pressure at sea level and temperature were taken in to account, i.e. the values used in Equation (4) were chosen to be 1000 to 1030 millibar, and 5 to 25°C, respectively.

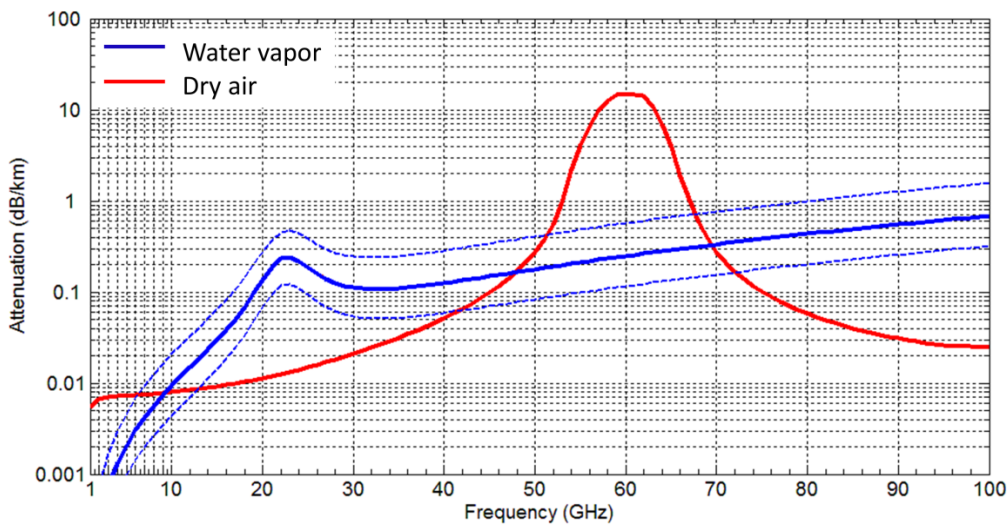

Figure 3.  Atmospheric gas attenuation. Signal loss as a result of water vapor (blue) and dry air (red).

The dominant factor in dry air attenuation at this frequency range is Oxygen, but variations in atmospheric $O_2$ concentrations take place at the ppm (parts per million) level, and therefore are negligible compared to the large 21% $O_2$ background (Keeling and Shertz, 1992). In order to estimate the effect of variations in atmospheric gas concentrations on measurements carried out for this paper, I focus on the frequency range around 40 GHz, which was used in this case (Measurements are presented in the results section). One can note that the typical fluctuation of attenuation as a result of atmospheric gases can reach $0.1-0.2$ dB/km (and, as mentioned, is due primarily to variations in water vapor concentration).

### 2.3 Calculating fog induced attenuation using real data measurements

Let us take a set of N links operating in the same frequency range over different path lengths $L_1, L_2 ... L_N$ (km), in a given area. It is assumed that the fog patch being observed is large enough to cover the whole set of N links measuring it simultaneously with conditions of average high relative humidity RH > 93% and measured by humidity gauges deployed in the observed area. Typically, fog is created when relative humidity is greater than 95% (e.g. Quan et al., 2011). However, since the humidity gauge used in this work has a measurement error of 2% at measurements above 90%, the selected moisture threshold is chosen to be 93%.

The occurrence of precipitation is ruled out using rain gauge measurements.

In order to determine the fog induced signal loss measured on each separate link, let us choose a reference set of $M$ measurements taken by each link close to the time of the event, against the RSL measurement taken on the foggy day itself – $RSL_{fi}$ (dBm). The attenuation for the *i-th* link - $\gamma_{fi}$ (dB) is calculated as follows:

$$(4) \qquad \gamma_{fi} = Median \ (RSL_{1i}, \ldots, RSL_{Mi}) - RSL_{fi} \quad (dB)$$

Notably, as a result of the typical conditions of high relative humidity, a thin layer of liquid water may condense on the microwave antenna radome, causing additional attenuation to that created by the fog that exists along the propagation path.

Under a linearity assumption the wet antenna signal loss- $\widehat{A_w}$ (dB) and the fog induced attenuation for unit of distance- $\widehat{\Gamma_f}$ (dB/km) are calculated from the measurements of all $i = 1, \ldots, N$ links:

$$(5) \qquad \widehat{\lambda}_f = \hat{\Gamma}_f \cdot L_i + \hat{A}_w \qquad (dB)$$

Where $\widehat{\lambda_f}$ (dB) is the estimated fog attenuation per given path length $L_i$.

Notably, $\widehat{A_w}$ (dB) denotes the calculated signal loss for an infinitesimal unit of distance and thus represents the attenuation caused by the possible condensation on the microwave antennas.

Sources of signal perturbations include variations in the refractive index of the atmosphere, white noise, quantization error of the microwave system. For further reading regarding detailed calculations of uncertainties, the reader is referred to David et al. (2013) and Zinevich et al. (2010).

# 3 RESULTS

The commercial microwave network, used in the events presented, operated in the frequency range between 37 and 39 GHz, and comprised 90-115 links for each event (based on the availability of data from the cellular provider in each case). The system stored RSL measurements with magnitude resolution of 0.1 dB for each link once every 24 hours at around 22:00 as reported in the quality of service reports of the cellular provider. Case 1 presents the obscureness of the fog from the satellite's perspective and Case 2 shows a situation in which fog is apparently detected from the satellite's point of view while in fact it is a low stratus cloud above ground level rather than fog. The conditions described in both cases continued for several hours per each event.

The reference RSL was determined for each link using the procedure described in equation (4) based on microwave measurements which were acquired during a 3 week period (between 3 to 24 November 2010, in accordance to the availability of data from the cellular provider). Meteorological reference data was received from Israel Meteorological Service (IMS) ground stations located in the observed area, and the satellite images taken by Meteosat Second Generation (MSG) were produced using CAPSAT (Clouds-Aerosols-Precipitation Satellite Analysis Tool (Lensky and Rosenfeld, 2008)).

## 3.1 Case I: The fog event of 15- 16 November, 2010

This fog event was reviewed in a previous paper where feasibility of the technique has been shown (David et al., 2013). In the current study this event is examined from a different vantage point where the ability to detect fog using the proposed technique when the satellite is limited, will be demonstrated.

Heavy fog, covering tens of square kilometers from the Sinai peninsula in Egypt, and along the Israeli coast and central plain (The Shfela) occurred from the early evening hours of the 15[th] till the morning hours of November 16[th], 2010 (David et al., 2013). At ground level, a Red Sea Trough (Goldreich, 2003) with a central axis moved Eastward, thus creating the conditions for Northwesterly flow from the Mediterranean Sea to enter the coastal region. At elevation, a deep ridge was moving Eastward. Measurements of IMS stations ruled out precipitation in the area during the event. Specifically, between 21:30 and 22:30 (i.e. around 22:00- the time when the measurements were taken by the microwave system) the relative humidity registered by the IMS ground stations in the test site was found to be between 93% and 97% with winds of ~1 to ~3 m/s – favorable conditions for the creation of fog.

Figure 4(a) shows a satellite image taken by Meteosat Second Generation (MSG) on November 16[th], 2010 at 00:29. The observed area is indicated by a rectangle. The figure reveals the large dimensions of the low stratus, depicted as a white shadow in the image. Figure 4(b) presents visibility estimates of two human observers, and Meteorological Optical Range (MOR) measurements of a sensor, that were located in the observed area and detected fog (i.e. visibility lower than 1 km[1]) between the hours of 21:50 and 07:00 between November 15th and 16th, 2010. The detected fog was heavy, and visibility dropped to a few tens of meters at times during the event.

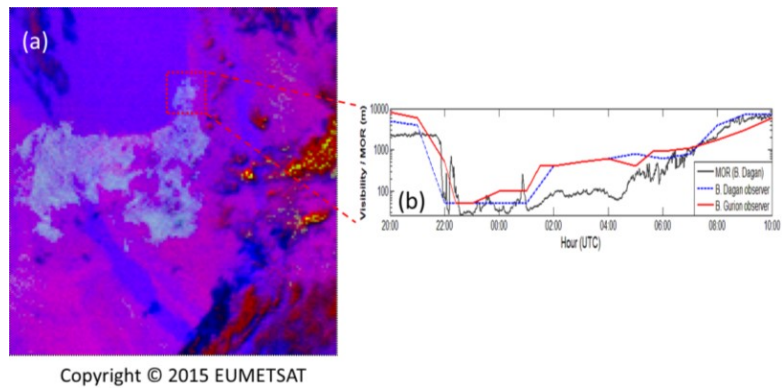

Figure 4. (a) Satellite image taken by Meteosat Second Generation (MSG) on 16 November 2010 at 00:29. The image was generated using the Clouds-Aerosols-Precipitation Satellite Analysis Tool and shows the large spatial coverage of the fog across the area (indicated by a white shadow). (b) Visibility estimations and Meteorological Optical Range (MOR) measurements taken during the fog event by the Beit Dagan human observer (red), the Ben Gurion human observer (black) and by the MOR sensor located at Beit Dagan (blue). Notably, heavy fog was detected by the different means starting from 21:50 on 15 November till the morning hours of the following day. Credit: figure 4(b) was regenerated using data from David et al. (2013).

Figure 5(a) shows a satellite image taken at 22:00 (UTC) on November 15th, of the event. While fog was detected in the area by the different ground based tools (as presented in Figure 4(b)), one can note that the phenomenon cannot be detected from the satellite view point due to high altitude cloud cover, shown in the image as a red shadow (Lensky and Rosenfeld, 2008), that obscured the ground level fog. Microwave links deployed in the area, on the other hand, measured increased attenuation during the same time due to the heavy regional fog. Figure 5(b) depicts the microwave system, which comprised 115 links deployed in the area, as a set of lines. The links shown as red lines detected attenuation in the area, while those in black did not measure any signal loss when compared to a day where no fog occurred, which was used as a reference measure.

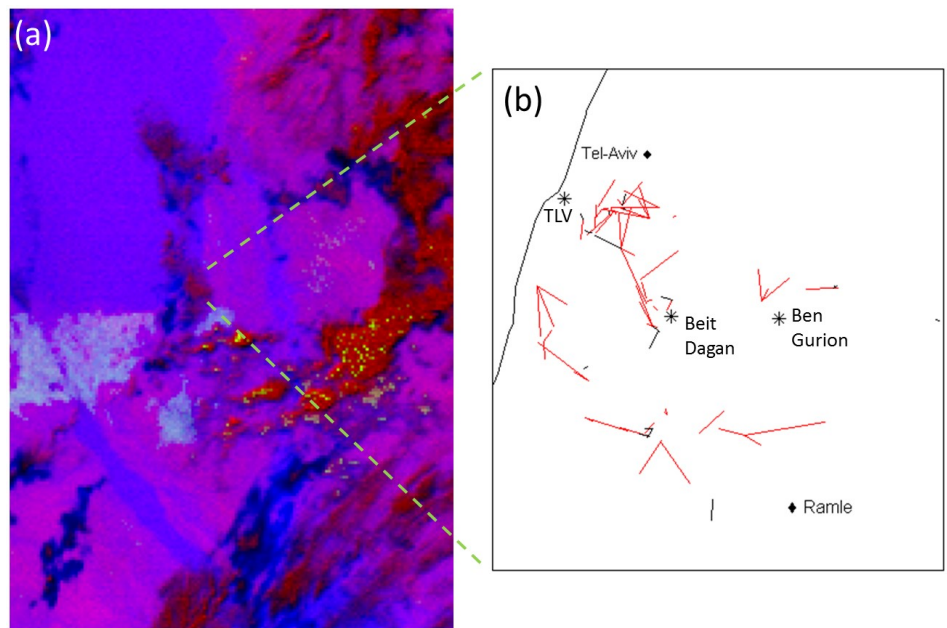

Figure 5. (a) Satellite image taken by MSG at 21:59 (generated using the CAPSAT). Notably, the areal fog patch was obscured by high level cloud layers denoted as a red shadow (the area is marked by the arrow). (b) Microwave link deployment across the test area. The measurements of the microwave system were taken at 22:00, i.e. approximately at the same time when the satellite observations (Fig. 3(a)) were acquired. The red lines indicate microwave links that measured excess attenuation due to fog. Black lines indicate locations of microwave links which were not measuring signal loss during the event. Meteorological stations are indicated by asterisks.

Figure 5(a) presents the attenuation measurements acquired during the fog event at 22:00 UTC on 15 November 2010 for each of the microwave links as a function of its length (red circles) vs. the measurements from the same link system on November 10th, 2010 (at 22:00, black circles) which was a humid day without fog according to the different ground based tools and the satellite.

The linear fits for the measurement sets taken on the foggy night and the night without fog are labeled (1) and (2) respectively in Figure 6. Notably, the slope of the linear fit, and the y-axis intercept for the foggy day are an order of magnitude greater than those calculated for the day without fog, where these two variables tend approximately to zero.

Figure 6(b) presents the same microwave system measurements from the foggy night. In accordance with figure 5(b), the red asterisks indicate measurements where fog induced attenuation was detected, where those in black indicate measurements where no fog induced attenuation was recorded.

The microwave system's differentiation between a foggy day and one without fog is apparent (Fig. 6(a)). Particularly noticeable is the system's ability to collect spatial information about the phenomenon, near ground level, while the satellite is limited in detecting fog due to high altitude cloud cover above the area (Fig. 5, Fig 6(b)).

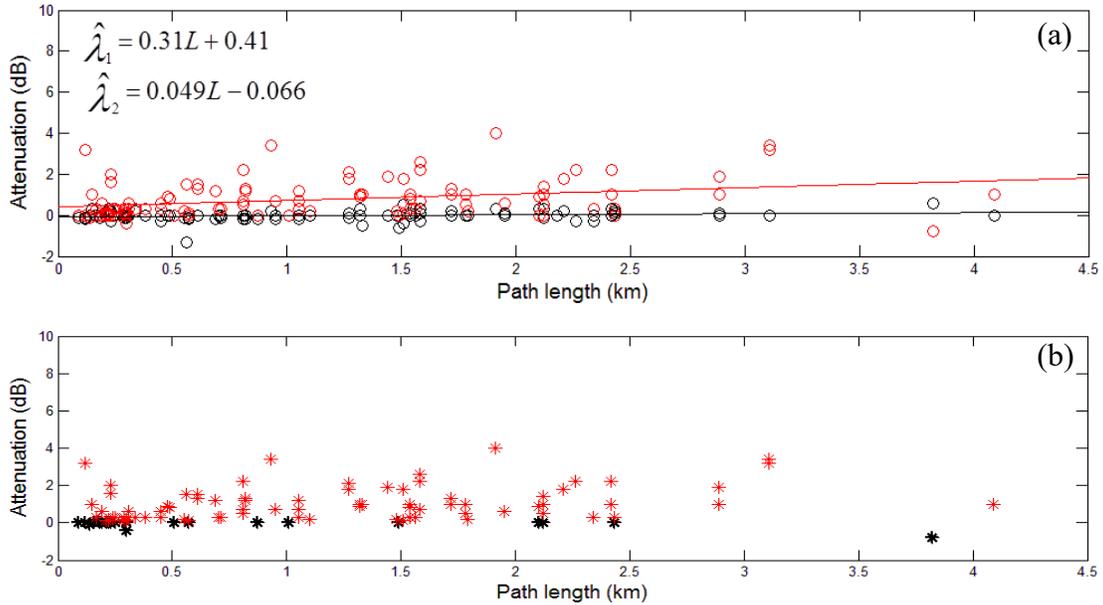

Figure 6. The microwave system measurements. (a) The measurements which were taken during the foggy night – November 15th, 2010 are denoted by the red circles and those acquired during the humid night – November 10th, 2010 are signified by the black circles. Each circle indicates a measurement form a single link, which were all taken simultaneously at 22:00. Linear fit approximations for the measurements from each night are given at the top of the panel. (b) The microwave system measurements from the foggy night of November 15[th], 2010, at 22:00 (This same set of measurements is shown as red circles in Figure 5(a)). Here, the red asterisks indicate added attenuation measured by each link, while the black ones indicate links that did not detect added attenuation on the foggy day, in accordance with the depiction in Figure 5(b).

### 3.2 Case II: The low stratus event of 11 November, 2010

On the night between November 11[th] and 12[th], 2010, a low stratus cloud covered vast areas of Israel's coastal and central plains. The phenomenon evolved through the combination of a shallow Red Sea Trough at ground level, with weak, disordered flows to the coast, and at elevation, in the central plain – a weak ridge.

Figure 7(a) shows the satellite image taken during the event (at 21:59 on November 11[th], 2010). The low stratus appears in the image as a white shadow. Note however, that it is impossible to identify, based on this satellite measurement, whether the prevailing conditions are fog, or a low lying stratus cloud, because of the limitations of the satellite in differentiating between these phenomena from its vantage point. Figure 7(b) shows the measurements of the microwave system deployed in the area (Figure 3b), near ground level (at elevations of tens of meters AGL), which was taking measurements at 22:00, i.e. with the greatest proximity to the satellite measurements. Particularly, the microwave system did not measure excess attenuation thus ruling out the possibility of heavy fog in the area. Figure 7(c) presents the visibility estimates received from the human observers and the MOR instrument which also ruled out the existence of fog at ground level. The visibility estimates measured by these means around 22:00 ranged from 3-5 km (human observers) to MOR of ~1.5 km according to the MOR instrument.

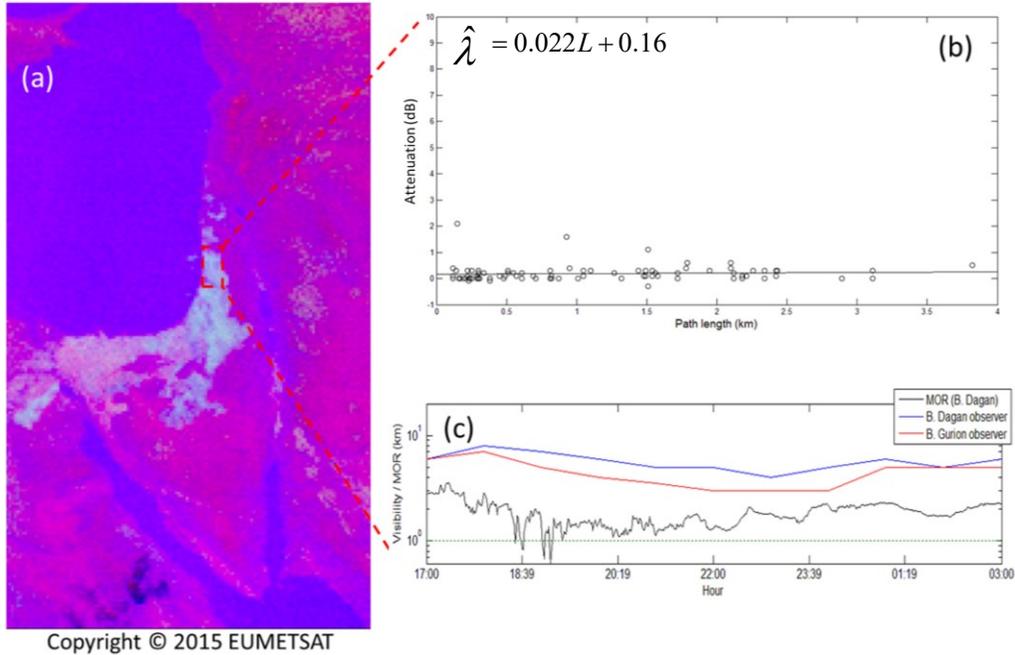

$$\hat{\lambda} = 0.022L + 0.16$$

Figure 7. The measurements of the different fog monitoring means. (a) The MSG measurement as taken on November 11[th], 2010 at 21:59 (generated using the CAPSAT). The low stratus observed appears as a white shadow. (b) The measurements of the microwave system deployed in the area. The figure is based on measurements from 90 links (the available measurements for that night from the cellular provider) out of the link map shown in Figure 4(b). (c) MOR estimates as taken by the sensor (black graph), the human observer in Beit Dagan (dashed blue line) and the observer in Ben Gurion Airport (red).

# 4 DISCUSSION

Satellite systems for fog monitoring have difficulties detecting the phenomenon during periods of high or medium altitude cloud cover due to the occlusions created. Further, the ability of such systems to differentiate between ground level fog and stratus clouds at higher elevations is limited. This work points to the potential of commercial microwave links as a complimentary solution to such challenging conditions.

Future research will need to investigate additional events from other regions with a goal of developing robust algorithms for automatically detecting the phenomenon given the newly available data proposed by this technology. Specifically, developing tools to combine near ground microwave measurements and satellite observations is clearly of interest, and may improve performance at minimal costs, since the commercial microwave infrastructure is already deployed and operating anyways.

There are a number of factors that can cause uncertainty in the measurements of the links, including: the accumulation of liquid water on the microwave antennas due to the high RH conditions typical of fog (Harel et al., 2015), uncertainty in the determination of the reference RSL according to which the fog induced attenuation is estimated (Zinevich et al., 2010), changes in humidity (David et al., 2009) and variations in the atmospheric refractive index (David and Gao, 2018). The ability to minimize the effects of these factors lies in the number of virtual sensors (links) that can provide a statistical indication of whether or not fog was present. In addition, the very selection of links that operate in the higher frequency bands (e.g. around 38 GHz in this work) increases the sensitivity to fog and deflects from the water vapor absorption line around 22.2 GHz (Figs. 1 and 2). Also, links that operate at higher frequencies are typically shorter and therefore their sensitivity to the effects of variations in the atmospheric refractive index is lower. Lack of synchronization, or differences between the clocks of the different measuring instruments (humidity gauges, visibility sensor, links, etc.) may introduce an additional uncertainty (however, since the conditions described in the presented events occurred over a period of several hours, this possible factor is negligible for the purpose of showing the feasibility of this study).

Moreover, the use of available meteorological tools in combination with the microwave links can negate other atmospheric phenomena (for example, ruling out rainfall using rain gauges and identification of high RH conditions using humidity gauge records, as done here). Having said that, the purpose of this study is to indicate the ability of the proposed system to provide an additional supplementary indication for detecting fog at times when satellite is not able to provide an answer, and further investigation will be required in future work in order to further test this ability.

## 5 CONCLUSIONS

The presented results demonstrate a potential tool that can provide observation in a manner that will assist end-users, such as operational forecasters, to reach a reliable, informed decision, in real-time (Chwala et al., 2016). As an example, one can consider a scenario where a fog front is approaching an airport, at a time when the satellite is limited in detecting it, due, for instance, to the reasons demonstrated here. Many airports are equipped with proprietary instruments for monitoring fog, however fog detection, in this case, would only occur once it has already reached the airport. The safety and economic implications of precise forecaster decision making regarding the issuing of a warning, prior to its actual occurrence, are meaningful.

It is important to note, that due to the increasing needs for high speed data transmission over the cellular network, there is a trend of moving to higher operating frequencies for these links, a development that may potentially allow for fog monitoring at higher resolution (David et al., 2015).

LWC in fog changes in space, altitude, and over time, and is dependent on surface and atmospheric conditions (Kunkel, 1984; Gultepe et al., 2007b). In Fig. 5(b), one can note that in a given area some links measured increased attenuation during the fog event, and some did not. While additional research on this topic is required, this observation may indicate the potential of the system, deployed over a wide spatial range, to allow for the detection of fog in certain areas, and ruling out the existence of fog in other areas.

## ACKNOWLEDGMENTS

I would like to express my deepest gratitude to my advisor Prof. H. Oliver Gao (Cornell University) for his fruitful advice and valuable discussions. I extend profound thanks to Prof. Pinhas Alpert and Prof. Hagit Messer Yaron (Tel Aviv University) for their continuous support in my academic pursuits. In addition, I wish to thank the Israeli Meteorological Service (IMS) and EUMETSAT for meteorological and satellite data. Special gratitude to the weather specialist Asaf Rayitsfeld (IMS) for the fruitful discussions. The microwave data were provided by Cellcom, Pelephone and PHI to the research team of Tel Aviv University to whom the author was affiliated. The author is grateful to N. Dvela, A. Hival and Y. Shachar (Pelephone); E. Levi, Y. Koriat, B. Bar and I. Alexandrovitz (Cellcom); Y. Bar Asher, O. Tzur, Y. Sebton, A. Polikar and O. Borukhov (PHI). The author acknowledges partial funding support by Cornell Atkinson Center for a Sustainable Future, National Science Foundation project CMMI-1462289, the Natural Science Foundation of China (NSFC) project # 71428001, and the Lloyd's Register Foundation, UK.